\documentclass[10pt,a4paper,final]{iopart}

\usepackage[dvips]{graphicx}
\usepackage{color}
\usepackage{ulem}

\usepackage{dcolumn}

\usepackage{bm}
\usepackage{amssymb}
\def\be{\begin{equation}}
\def\ee{\end{equation}}
\def\bea{\begin{eqnarray}}
\def\eea{\end{eqnarray}}

\def\be{\begin{equation}}
\def\ee{\end{equation}}

\begin{document}

\title[The influence of anisotropic Rashba spin-orbit coupling on current-induced spin polarization in graphene]{The influence of anisotropic Rashba spin-orbit coupling on current-induced spin polarization in graphene}

\author{Mir Vahid Hosseini}
\address{Department of Physics, Faculty of Science, University of Zanjan, Zanjan 45371-38791, Iran}
\ead{mv.hosseini@znu.ac.ir}

\begin{abstract}
We consider a disordered graphene layer with anisotropic Rashba spin-orbit coupling subjected to a longitudinal electric field. Using the linear response theory we calculate current-induced spin polarization including in-plane normal and parallel components with respect to the electric field direction. Unlike the case of isotropic Rashba spin-orbit where the normal component of spin polarization is linear in terms of Fermi energy around the Dirac point, anisotropic Rashba spin-orbit can result in non-linear dependence of this component at such energies within the Lifshitz points. Furthermore, we show that anisotropic Rashba interaction allows for tuning the direction of spin polarization from perpendicular direction to the parallel one such that for certain values of Rashba parameters the magnitudes of both components can also be quenched. The effect of carriers scattering on randomly distributed non-magnetic disorders is also taken into account by calculating vertex correction. This results in modification of spin polarization components depending on the relative strength of Rashba parameters.
\end{abstract}

\maketitle


\section {Introduction} \label{s1}
Manipulation of spin degree of freedom by means of an externally applied electric field \cite{CIPRev3,CIPRev4,CIPRev5} is one of the exciting phenomena in the field of spintronics. It also has attracted a great interest from both theoretical \cite{TheCIP} and practical \cite{ExpCIP} aspects. To manipulate the electron spin, spin-orbit interaction, which couples an electron spin to its momentum \cite{Winkler1}, plays a key factor \cite{Dyakonov}, in addition to the some other effects \cite{OtherEff1,OtherEff2}. Various effects have been predicted and identified so far to generate spin-orbit interaction depending on the type and structure of materials \cite{SODuine}. For instance, Rashba \cite{Rashba1} and Dresselhaus \cite{Dressel} spin-orbit interactions, respectively, arise from bulk inversion asymmetry and structural inversion asymmetry. Whereas some specific features of material may give rise to intrinsic spin-orbit coupling.

Beside silicon-based compounds, one of the promising candidates for future electronic devices is two-dimensional allotrope of carbon atoms arranged in hexagonal lattice which is known as graphene \cite{graphene}. Despite smallness of spin-orbit interaction \cite{SOIGRaphene} in pristine graphene, recently, intensive efforts have been done to enhance this effect in graphene \cite{SOenhanceGraph1,SOenhanceGraph3,SOenhanceGraph4,SOenhanceGraph5}. Very recently, it has been observed experimentally that in graphene grown by chemical vapour deposition on ​copper \cite{SOcop} and on Ir(111) substrate with intercalated Pt monolayer \cite{SOpt}, anisotropic Rashba spin-orbit coupling can be induced through proximity effect. This opens up new prospects for many potential applications. In this regards, to control charge and spin of carriers, the interaction of spin-orbit-coupled graphene layer with externally applied fields can exhibit some spectacular characteristics \cite{graphRev1,graphRev3}.

The presence of an electric current along with spin-orbit coupling can induce a non-zero average spin polarization in charge carriers when flowing through a non-magnetic system. This current-induced spin polarization, which is very distinct from spin Hall effect \cite{CIPRev3,CIPRev4,CIPRev5,SpinHE}, is known as Edelstein \cite{Edelstein} or inverse spin-galvanic effect \cite{InSpGal1,InSpGal3,InSpGal4}. Edelstein effect results from asymmetric spin relaxation processes \cite{CIPRev3,CIPRev4,CIPRev5,Edelstein,InSpGal1,InSpGal3,InSpGal4}. The current-induced spin polarization has a polarization vector whose orientation is dependent on the point group symmetry of structure and the relative strengths of the structure and bulk inversion asymmetries \cite{SBInvSym2,AnisoCIP}.

In the previous studies, electronic transport through homogeneous \cite{homoTrans1,homoTrans2,homoTrans3,homoTrans6,homoTrans7} and inhomogeneous \cite{InhomoHoss1,InhomoHoss2,InhomoHoss3} Rashba spin-orbit region in graphene has been discussed. Also, the study of current-induced spin polarization in various systems including electron and hole carriers has been performed by some previous works theoretically \cite{OtherEff1,AnisoCIP,CurrIndgeneraTheo2,CurrIndgeneraTheo3,CurrIndgeneraTheo4,CurrIndgeneraTheo6,CurrIndgeneraTheo8,CurrIndgeneraTheo9,CurrIndgeneraTheo11,CurrIndgeneraTheo14,CurrIndgeneraTheo16,CurrIndgeneraTheo17} and experimentally \cite{ExpCIP,CurrIndgeneraExper2,CurrIndgeneraExper3,CurrIndgeneraExper5,CurrIndgeneraExper7}, (see also, e.g., Ref. \cite{InSpGal1,InSpGal3,InSpGal4} and references therein). But it is still interesting to investigate the Edelstein effect in materials with unique features such as hybrid heterostructures \cite{CurrIndGraph-TI1,CurrIndGraph-TI2}, carbon nanotubes \cite{CurrIndCarNanoTub}, and topological insulators \cite{CurrIndTI1,CurrIndTI2,CurrIndTI3,CurrIndTI4,CurrIndTI5,CurrIndTIRoomTep}. In addition, supercurrent-induced spin polarization in superconducting heterostructures comprised of conventional materials \cite{SupCurConv1,SupCurConv2,SupCurConv3} and topological insulators \cite{SupCurTI} has been demonstrated.

Recently, the properties of spin-polarized current via electric field \cite{curSpinGraph1,curSpinGraph2} and photoinduced spin-current \cite{photoSpinGraph1,photoSpinGraph2} have been considered in graphene layer with isotropic Rashba spin-orbit interaction. In the former case, it has also been shown the coupling between charge current and spin polarization due to adatom impurities \cite{AnisoRashGraph3}. Furthermore, the effect of anisotropic Rashba spin-orbit interaction on current-induced spin polarization in conventional material has been revealed theoretically \cite{TheoAniso} and experimentally \cite{ExpAniso}. Also, it was pointed out that, within the Boltzmann equation approach, the effects of Rashba anisotropy \cite{AnisoRashGraph1} can strongly modify the spin Hall angle in graphene \cite{AnisoRashGraph2}.

In graphene, however, the effect of anisotropic Rashba spin-orbit coupling on the current-induced spin polarization lacks a detailed study. Contrary to traditional two-dimensional materials, in graphene, Rashba Hamiltonian is momentum independent at low energies stemming from triangular symmetry of the lattice structure. Correspondingly, spin degeneracy can be lifted by coupling between spin and pseudospin. So, anisotropic Rashba spin-orbit coupling, having different weights of Rashba parameters, may provide exciting features, such as unusual asymmetric spin relaxation. It would be interesting if anisotropic Rashba spin-orbit coupling could be served as a practical tool for manipulating both magnitude and direction of spin polarization efficiently which is absent in the isotropic case.

In this paper, we address the properties of current-induced spin polarization in a graphene layer focusing on the role of anisotropic Rashba spin-orbit interaction. In the present case, similar to the case of two-dimensional electron gas \cite{TheoAniso} anisotropic Rashba coupling provides Lifshitz points in the conduction and valance bands which are located at $K$ points of the Brillouin zone. Using Green's function method and linear response theory we argue that the in-plane spin-polarized current, in addition to the spin polarization normal to current direction, has a component along the current direction. Furthermore, we find that the spin polarization perpendicular to the current at low energies, within Lifshitz points, reveals a non-linear behavior as a function of chemical potential which is in contrast to the isotropic case \cite{curSpinGraph1}. Vertex correction, owing to scattering on non-magnetic disorders, is also incorporated. Depending on the strengths of Rashba spin-orbit parameters in different directions, vertex correction affects the strength of scattered spin components differently.

The outline of the paper is as follows. In Sec. \ref{s2}, we introduce the model of Hamiltonian with anisotropic Rashba coupling and present the theory of current-induced spin polarization in the linear response regime. In Sec. \ref{s3}, we present the results of spin polarization obtained from analytical derivation, and explain the observed results by numerics. In Sec. \ref{s4}, disorder scattering effect on current-induced spin polarization is investigated by including vertex correction. Finally, concluding remarks are discussed in Sec. \ref{s5}.
\begin{figure}
 \centering
  \includegraphics[width=6cm]{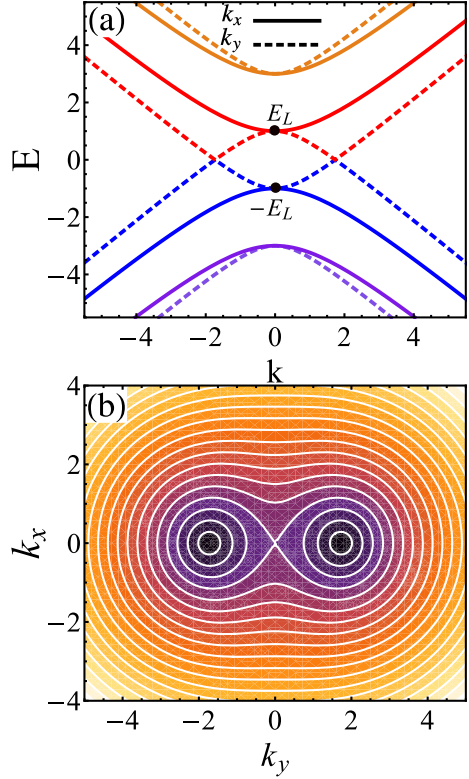}
  \caption{(Color online) (a) Sketch of the energy band structure of a graphene layer with anisotropic Rashba spin-orbit coupling. Solid
lines indicate energy states along $k_x$, while dashed lines denote the energy states along the perpendicular direction, $k_y$. (b) Contour plot of constant energy lines of lowest conduction energy band. Here, $\lambda_2 = 2\lambda_1 = 2$.}
  \label{fig1}
\end{figure}
\section {Model and Theory}\label{s2}
We consider a graphene layer on a substrate with impurities, which induces anisotropic Rashba coupling in graphene. The resulting disorder potentials are assumed to be smooth enough as compared with the atomic scale of graphene so that intervalley scattering can be ignored. Thus, we will restrict ourself to the single valley picture and incorporate the effect of disorders via vertex correction. The low energy excitation of electrons in graphene including anisotropic Rashba interaction is well described by the Hamiltonian \cite{AnisoRashGraph1,AnisoRashGraph2}
\begin{equation}
H = \hbar v_f(k_x\tau_x + k_y\tau_y)\otimes\sigma_0 + \lambda_1\tau_x\otimes\sigma_y - \lambda_2\tau_y\otimes\sigma_x,
\label{Hamil}
\end{equation}
where the first term is two-dimensional Dirac Hamiltonian, the last two terms are anisotropic Rashba Hamiltonian, $v_f$ is the Fermi velocity, $\lambda_{1,2}$ are the strengths of Rashba coupling. Also, $\boldsymbol \sigma$ and $\boldsymbol{\tau}$ are the Pauli vectors acting on the spin and pseudospin degrees of freedom, respectively. $\sigma_{0}$ is a unit matrix in the spin space. For convenience, hereafter we choose the lattice constant $a = 1$ as the length unit and $\hbar v_f = 1$ as the energy unit. Due to the reduced orthorhombic symmetry of the graphene, the two terms of Rashba coupling have different weight, originating from the breakdown of both mirror and sixfold rotation symmetry. Here, we have assumed that Rashba coupling strengths are position-independent and also the effect of intrinsic spin-orbit coupling is negligible. The eigenvalues can be easily obtained by diagonalazaing the model Hamiltonian (\ref{Hamil}) which are given by,
\begin{eqnarray}\label{Eigenvalues}
E_{1,2}& = &\sqrt{k^2_x + k^2_y + \lambda^2_1 + \lambda^2_2\pm2\xi}, \\
E_{3,4}& = & -\sqrt{k^2_x + k^2_y + \lambda^2_1 + \lambda^2_2\mp2\xi},
\end{eqnarray}
where $E_{1,2} (E_{3,4})$ are conduction (valance) bands with $\xi = \sqrt{(k_x\lambda_1)^2+(k_y\lambda_2)^2+(\lambda_1\lambda_2)^2}$. Clearly, one may expect that the band spectra are anisotropic because of anisotropic Rashba spin-orbit coupling. In Fig. \ref{fig1}(a), the energy band structure of the system is presented along $k_x$ and $k_y$. For $|\mathbf{k}| \ll 1$, while both upper conduction and lower valance bands are isotropic, the other two bands are anisotropic. But away from Dirac point, almost same band anisotropy appears for all the bands. Unlike the energy spectrum of graphene with isotropic Rashba spin-orbit coupling, band inversion takes place in the lowest-energy conduction band and highest-energy valence band so that these bands are degenerate at two points $\pm\sqrt{\lambda^2_2 - \lambda^2_1}$, located symmetrically around the $\boldsymbol K$ point along the $k_y$ for $|\lambda_2| > |\lambda_1|$. For opposite case, i.e., $|\lambda_2| < |\lambda_1|$, the two degenerate points take place at $\pm\sqrt{\lambda^2_1 - \lambda^2_2}$ in the $k_x$ direction. The band inversion due to anisotropic spin orbit coupling leads to changing of Fermi line topology from a closed curve to two disconnected closed curves resulting in a Lifshitz transition \cite{lifshtz} as shown in Fig. \ref{fig1}(b). The Lifshitz transition point comprising of a single Van Hove singularity occurs at energies
\begin{equation}
\pm E_L = \pm||\lambda_1| - |\lambda_2||,
\label{Lif}
\end{equation}
where $+$ ($-$) indicates the Lifshitz point in the conduction (valance) band.

\begin{figure}
 \centering
  \includegraphics[width=6cm]{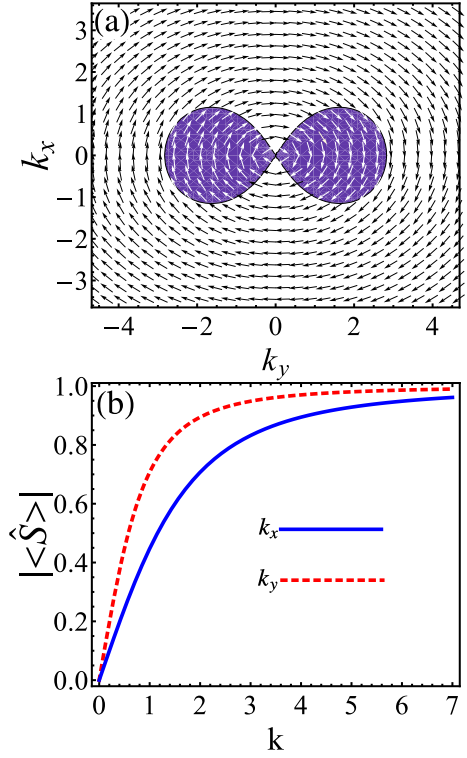}
  \caption{(Color online) (a) Spin orientation of states below (dark region) and above (white region) Lifshitz point of lowest conduction band. (b) The magnitude of expectation value of electron spin versus wave vector $k$. Solid (dashed) line indicates expectation value of electron spin along $k_x$ ($k_y$). Here, $\lambda_2 = 2\lambda_1 = 2$.}
  \label{fig2}
\end{figure}

The expectation value of electron spin $\langle\hat{\mathbf{S}}\rangle$, exhibiting the spin orientation, for conduction bands $E_1$ and $E_2$ are
\begin{eqnarray}\label{expSpin}
\langle\hat{\mathbf{S}}\rangle_{1,2} = \frac{\mp k_y\lambda_2\hat{k}_x\pm k_x\lambda_1\hat{k}_y}{\xi},
\end{eqnarray}
where the unit vector $\hat{k}_{x(y)}$ is along the $k_{x(y)}$ direction. Here, for the case of graphene we have defined
\begin{equation}
\hat{S}_{\alpha} = \frac{\hbar}{2}\tau_{0}\otimes\sigma_{\alpha},
\end{equation}%
where $\tau_{0}$ is a unit matrix in the pseudospin space.
Equation (\ref{expSpin}) indicates that similarly as conventional Rashba spin-orbit interaction, the magnitudes and orientations of the expectation values of electron spin respectively are identical and opposite in bands $E_1$ and $E_2$. Moreover, the plane $E = 0$ acts as a mirror for the electron spin orientation in the valance bands. From Eq. (\ref{expSpin}), on the other hand, one can see that the spin alignment is not always perpendicular to the direction of $\mathbf{k}$ which is in contrast to the conventional Rashba spin-orbit coupling case. The orientation of spin states for the lowest conduction band is represented in Fig. \ref{fig2} (a). The states with energies below the Lifshitz point are shown in blue (dark) region, while those for which white area is attributed have energies larger than the Lifshitz point energy. Also, the magnitude of spin expectation value, $|\langle\hat{\mathbf{S}}\rangle|$, saturates at large $k$ but it is not isotropic, as shown in Fig. \ref{fig2} (b). The saturation is achieved faster along a direction where the band inversion occurs. These features affect spin polarization imposed by external electric field as will be discussed below.

According to the linear response theory, the response of electron spin polarization $\mathbf{S}$ to externally applied electric field $\mathbf{E}$ is defined as,
\begin{equation}
S_{\alpha} = \sum_{\beta}\chi_{\alpha\beta}E_{\beta},
\end{equation}
where $\chi_{\alpha\beta}$ is the electric spin susceptibility. By taking into account vertex correction \cite{Edelstein,SpiSuscVertexCorr}, $\chi_{\alpha\beta}$ can be calculated by Kubo formula \cite{SpiSusc1},
\begin{equation}
\chi_{\alpha\beta} = -\frac{e\hbar }{2\pi }\int dE\frac{\partial f(E)}{%
\partial E}Tr\langle \hat{S}_{\alpha}G^{R}(E)v_{\beta}G^{A}(E)\rangle _{c},  \label{Kubo2}
\end{equation}%
where $G^{R(A)}_{\mathbf{k}}(E)$ is the retarded (advanced) Green's function associated with Eq. (\ref{Hamil}), $f(E)$ is the
Fermi distribution function, $v_{\beta} = (\partial H/\partial k_{\beta})/\hbar$ is the velocity matrix along the $\beta$-th axis, and the bracket $\langle \cdots \rangle _{c}$ represents the ensemble averaging over impurity configuration. $Tr$ means trace over pseudospin and spin spaces. In the ladder approximation, which is valid for low-density Fermi system, Eq. (\ref{Kubo2}) for $\chi_{\alpha\beta}$ at zero temperature can be written as
\begin{equation}
\chi_{\alpha\beta} = \frac{e\hbar}{2\pi} \int \frac{d^{2}k}{(2\pi )^{2}}Tr\left[ \tilde{S}%
_{\alpha}G^{R}_{\mathbf{k}}(\mu) v_{\beta} G^{A}_{\mathbf{k}}(\mu)\right] ,  \label{Kubo3}
\end{equation}%
where $\mu$ is the chemical potential measured from Dirac point and the spin-vertex function $\tilde{S}_{\alpha}$ satisfies the self-consistent equation~\cite{SpiSusc1}
\begin{equation}
\tilde{S}_{\alpha} = \hat{S}_{\alpha} + n_{imp}V_{0}^{2}\int \frac{d^{2}k}{(2\pi
)^{2}}G^{A}_\mathbf{k}(\mu)\tilde{S}_{\alpha}G^{R}_\mathbf{k}(\mu),
\label{VereqnkR3}
\end{equation}%
where $n_{imp}$ is the impurity density. $V_{0}$ is the strength of scattering potential of identical pointlike impurities distributed randomly in the form
\begin{equation}
V(\mathbf{r}) = V_{0}\tau_0\otimes\sigma_0\sum_{i}\delta(\mathbf{r}-\mathbf{R}_{i}).
\end{equation}%
Here, we have assumed that the scattering potential conserves both the spin and pseudospin states in the scattering process. This assumption implies that nonmagnetic impurities are distributed equally on two different sublattices. Choosing the electric field to be along the $y$ direction and using the Green's function based on Eq. (\ref{Hamil}), we find that non-zero electric spin susceptibilities without vertex correction are,
\begin{equation}
\chi _{xy(yy)} =\frac{e\hbar^2}{(2\pi)^3} \int dkd\theta \frac{g_{1(2)}(k,\theta)}{\prod_{n=1}^4[(\mu-E_n)^2+\Gamma^2]} ,  \label{spxy}
\end{equation}%
with
\begin{eqnarray}
g_1(k,\theta) = \!&8&\!\!k\lambda_2\mu [k^4-2k^2(\lambda^2_2-\lambda^2_1)-(\lambda^2_1-\lambda^2_2)^2-\mu^4\nonumber\\
&+&2\mu^2(\lambda^2_1+\lambda^2_2)-2k^2(k^2-\mu^2)\cos(2\theta)], \label{cofxy}
\end{eqnarray}%
and
\begin{figure}
\centering
\includegraphics[width=6cm]{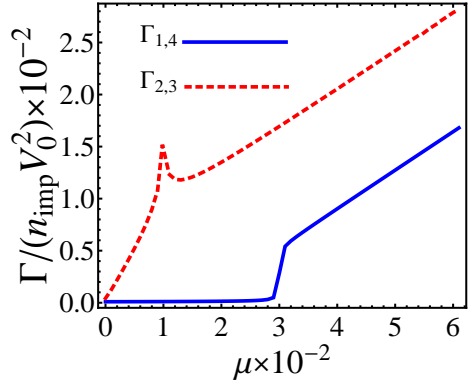}\\
\caption{(Color online) (a) Relaxation rates $\Gamma$ of quasiparticles in different energy bands of a graphene layer with anisotropic Rashba spin-orbit coupling versus $\mu$. Solid (dashed) line shows $\Gamma_1$ and $\Gamma_4$ ($\Gamma_2$ and $\Gamma_3$) which have the same values. For $|\mu|$ larger than both $\lambda_1$ and $\lambda_2$ all the $\Gamma$'s change identical. Here, $\lambda_2 = 2\lambda_1 = 2\times10^{-2}$.}
\label{fig3}
\end{figure}
\begin{equation}
g_2(k,\theta) = -16k^3\lambda_1\mu(k^2+\lambda^2_1-\lambda^2_2-\mu^2)\sin(2\theta), \label{cofyy}
\end{equation}%
where $\theta = \arctan(k_y/k_x)$, $k = \sqrt{k^2_x + k^2_y}$ and dimensionless relaxation rate $\Gamma = -Im \Sigma_0 = \frac{a}{2v_f\tau}$ with $\Sigma_0$ and $\tau$ being self-energy and momentum-relaxation time, respectively. The imaginary part of self-energy originating from scattering of quasiparticles on the pointlike impurities is given by \cite{selfEnergy}
\begin{equation}
\Gamma_n = Im \Sigma_0 = n_{imp}V^2_0 \sum_{\mathbf{k}\in 1BZ}\delta(|\mu| - E_n), \label{gamm}
\end{equation}%
where $n = 1, 2, 3$ and $4$ are the band index. Equation (\ref{gamm}) implies that the quantities $\mu$, $n_{imp}V^2_0$, and $\Gamma$'s are related to each other so that by specifying two of them the third one will be determined. Using full band structure of graphene \cite{graphene} and ignoring trigonal warping effect, we evaluate numerically the summation of Eq. (\ref{gamm}) in the first Brillouin zone at low energies. In Fig. \ref{fig3} the dependence of $\Gamma$'s on $\mu$ is depicted for $\lambda_2 = 2 \lambda_1 = 2\times10^{-2}$. The relaxation rate $\Gamma_1 = \Gamma_4$ and $\Gamma_2 = \Gamma_3$, which results from particle-hole symmetry of the system Hamiltonian. For $|\mu| \ll \lambda_1,\lambda_2$ the values of $\Gamma_{2,3}$ increase non-linearly but $\Gamma_{1,4}$ acquire vanishingly small values, while for $|\mu| \gg \lambda_1,\lambda_2$ all the $\Gamma$'s increase linearly. There are states at which the diagrams of $\Gamma_{2,3}$ exhibit sharp peaks representing Lifshitz points.

In obtaining the results (\ref{spxy}), for concreteness, we have treated $\Gamma$'s as identical quantities. For $\Gamma$'s, in our numerical considerations, we take the average value $\Gamma = 10^{-4}$, unless otherwise stated. Notice that the real parts of self-energy renormalize the chemical potential and the Rashba parameters that can be ignored in the dilute impurity limit. Moreover, we have taken into account impurities effect within the Born approximation by including a finite time for relaxation of the quasiparticles and vertex corrections. The vertex corrections will be analyzed in Sec. \ref{s4}.

It is worthwhile mentioning that only when $\lambda_1 \neq \lambda_2$ due to anisotropy of $E_n$'s, the integration over $\theta$ does not vanish in $\chi _{yy}$. This behavior is different from that of usual two-dimensional cases \cite{Edelstein,curSpinGraph1}, where the in-plane polarization vector is perpendicular to the electric current. As a consequence, the spin polarization has not only perpendicular component but also parallel one with respect to the direction of electric current. The existence of spin polarization along the current direction can be traced back to the particular electron spin texture [see Eq. (\ref{expSpin})] in which the spin direction is not perpendicular to the momentum at some states. On the contrary, if $\lambda_1 = \lambda_2$ our results coincide with those reported in the previous studies \cite{curSpinGraph1}. Furthermore, from Eqs. (\ref{cofxy}) and (\ref{cofyy}) it can be seen that $\lambda_2$ and $\lambda_1$ have a direct impact on spin polarizations in the $x$ and $y$ directions, respectively, in our chosen basis for the spin space of Rashba Hamiltonian of Eq. (\ref{Hamil}).

\section {Analytical and Numerical results} \label{s3}
In this section, we analyze the effect of anisotropic Rashba spin-orbit coupling on current-induced spin polarization by making further analytical progress in some limiting cases and numerical calculation. For anisotropic Rashba couplings one can write the strengths of Rashba parameters as $\lambda_{1,2} = \lambda \pm \delta\lambda$ in terms of mean value $\lambda$ and their deviation $\delta\lambda$ from isotropic case. For small anisotropy, i.e., $|\delta\lambda| \ll 1$, up to the linear terms in $\delta\lambda$, for $\chi _{xy}$ the following analytical relation is obtained by using Cauchy's residue theorem,
\begin{equation}
\chi _{xy} = \frac{e\hbar^2}{8\pi\Gamma} \frac{(2\lambda\pm\mu)}{\lambda\pm\mu}[\mu\pm\frac{(2\lambda\mp\mu)}{\lambda}\delta\lambda] ,  \label{spxyexpand}
\end{equation}%
for the case $|\mu| < 2 \lambda$ and,
\begin{equation}
\chi _{xy} = \pm\frac{e\hbar^2}{4\pi\Gamma} \frac{\mu^2\lambda+(\mu^2-4\lambda^2)\delta\lambda}{\mu^2-\lambda^2} ,  \label{spyyexpand}
\end{equation}%
for the case $|\mu| > 2 \lambda$. Here the upper (lower) sign is for $\mu > 0$ ($\mu < 0$). Note also that Eqs. (\ref{spxyexpand}) and (\ref{spyyexpand}) have been derived in the limit of $\Gamma \rightarrow 0$. For $|\mu| \ll \lambda$, we find
\begin{equation}
\chi _{xy} = \frac{e\hbar^2}{4\pi\Gamma} (\mu\pm2\delta\lambda) ,  \label{spxyexpand0}
\end{equation}%
indicating that spin susceptibility $\chi_{xy}$ increases linearly as a function of $\mu$ similarly as isotropic Rashba coupling case in graphene. But the absolute value of $\chi_{xy}$ is either larger or smaller than that of isotropic case depending on the sign of $\delta\lambda$.
For the case $|\mu| \gg \lambda$, Eq. (\ref{spyyexpand}) reduces to
\begin{equation}
\chi _{xy} = \pm \frac{e\hbar^2}{4 \pi\Gamma} (\lambda + \delta\lambda) ,  \label{spyyexpand0}
\end{equation}%
denoting that the value of $\chi _{xy}$ saturates at large chemical potential with some deviations as compared to the isotropic case. Note that the above results are obtained for the case $\lambda > \delta\lambda$. On the other hand, in the opposite case, i.e., $\lambda < \delta\lambda$, the overall sign will be changed. However, even up to second order in $\delta\lambda$, $\chi _{yy}$ vanishes, therefore, we leave it for more detailed investigations with numerical calculations.
\begin{figure}
\centering
\includegraphics[width=6cm]{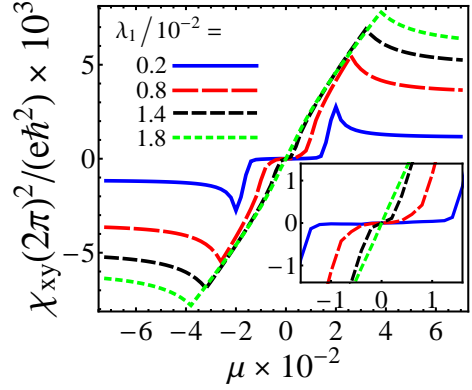}\\
\caption{(Color online) Dependence of $\chi _{xy}$ on $\mu$ for different values of Rashba spin-orbit coupling $\lambda_1$. Inset panel shows $\mu$ dependence of $\chi _{xy}$ at low energies. Here, $\lambda_2 = 1.8\times10^{-2}$.}
\label{fig4}
\end{figure}

Now, taking into account the band inversion discussed above we should numerically calculate the current-induced spin polarization in the $x$ and $y$ directions. In Fig. \ref{fig4}, the spin susceptibility $\chi _{xy}$ is shown as a function of chemical potential for different values $\lambda_1$ with $\lambda_2 = 1.8\times10^{-2}$. One can see that $\chi _{xy}$ increases in magnitude with increasing $\mu$ until reaches a maximum then with further increases the chemical potential it slightly decreases and finally saturates. In the inset panel, a close view around the Dirac point is shown. At such energy range, for unequal values of $\lambda_1$ and $\lambda_2$, $\chi _{xy}$ behaves non-linearly, whereas, for $\lambda_1 = \lambda_2$, it changes linearly. Note also that there is the same behavior for $\chi _{xy}$ versus $\mu$ for different values of $\lambda_2$ and fixed value of $\lambda_1$ which is not shown.

\begin{figure}
 \centering
 \includegraphics[width=8.5cm]{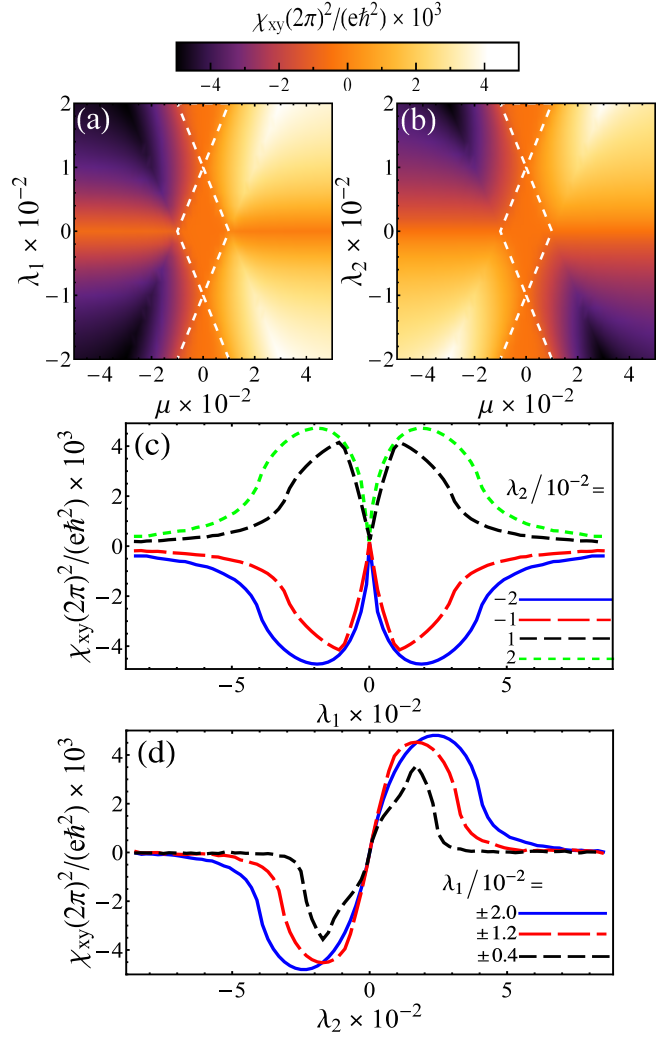}
  \caption{(Color online) Density plot of $\chi_{xy}$ as functions of $\mu$ and (a) [(b)] $\lambda_1$ ($\lambda_2$) with $\lambda_2 = 10^{-2}$ ($\lambda_1 = 10^{-2}$). Dashed lines indicate the energy states at which Lifshitz transition occurs. Also, plot of $\chi_{xy}$ as a function of (c) [(d)] $\lambda_1$ ($\lambda_2$) for various values of $\lambda_2$ ($\lambda_1$) with $\mu = 2\times10^{-2}$.}
  \label{fig5}
\end{figure}

In order to clarify that the above-described behaviors belong to which energy range, we illustrate the density plot of $\chi _{xy}$ as functions of $\mu$ and $\lambda_{1(2)}$ for $\lambda_{2(1)} = 10^{-2}$ in Fig. \ref{fig5}(a) [\ref{fig5}(b)]. From both plots, one can see that $\chi_{xy}$ is antisymmetric versus $\mu$ and $\lambda_2$ but it is symmetric as a function of $\lambda_1$. Using Eq. \ref{Lif}, the locations of Lifshitz points are also shown in the parameter space of Figs. \ref{fig5}(a) and \ref{fig5}(b) by dashed lines. Within the energies between the Lifshitz points of conduction and valance bands, i.e., $-E_L < \mu < E_L$, the absolute value of $\chi _{xy}$ increases non-linearly, while outside of this energy range the spin susceptibility reveals the same behavior as the isotropic Rashba coupling case. The non-linear dependence of $\chi_{xy}$ on chemical potential within the Lifshitz points is a result of low available spin state in such energies contributing to the scattering process (see Fig. \ref{fig2}(a)). Accordingly the spin susceptibility becomes weaker than that of isotropic case. For large $k$ more states are available such that beyond the Lifshitz points, the effect of band anisotropy is less dominant and available states become considerable. Subsequently, the results are consistent with those of isotropic case. Figure \ref{fig5}(c) [\ref{fig5}(d)] illustrates the dependence of $\chi _{xy}$ on $\lambda_1$ ($\lambda_2$) for different values of $\lambda_2$ ($\lambda_1$) with $\mu = 2\times10^{-2}$. From Fig. \ref{fig5}(c) one can see that for $\lambda_2 > 0$ ($\lambda_2 < 0$), $\chi _{xy}$ is positive (negative). However, the values of $\chi _{xy}$ are the same for both positive and negative values of $\lambda_1$ as shown in Fig. \ref{fig5}(d). Furthermore, in both Figs. \ref{fig5}(c) and \ref{fig5}(d), $\chi _{xy}$ increases in magnitude with increasing the absolute value of Rashba parameters and after passing a maximum decreases such that it finally tends to vanish. As a result, if the anisotropy of Rashba coupling becomes large enough, then spin polarization perpendicular to the current direction would be suppressed.

\begin{figure}
 \centering
 \includegraphics[width=8.5cm]{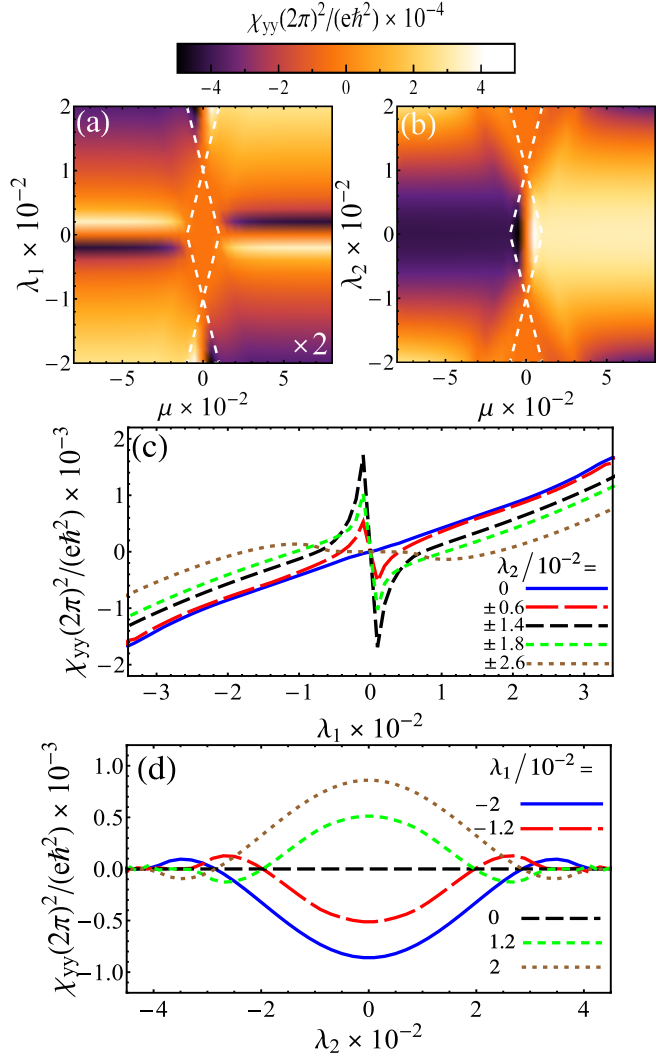}
  \caption{(Color online) Density plot of $\chi_{yy}$ as functions of $\mu$ and (a) [(b)] $\lambda_1$ ($\lambda_2$) with $\lambda_2 = 10^{-2}$ ($\lambda_1 = 10^{-2}$). Dashed lines indicate the energy states at which Lifshitz transition occurs. Also, plot of $\chi_{yy}$ as a function of (c) [(d)] $\lambda_1$ ($\lambda_2$) for various values of $\lambda_2$ ($\lambda_1$) with $\mu = 2\times10^{-2}$.}
  \label{fig6}
\end{figure}

In Figs. \ref{fig6}(a) and \ref{fig6}(b), the density plots of $\chi_{yy}$ are shown versus chemical potential and Rashba parameters. In both figures, for energies lying within Lifshitz points, i.e., states between dashed lines, if $|\lambda_1| > |\lambda_2|$, $\chi_{yy}$ changes its sign abruptly at $\mu = 0$ and it can reach largest magnitude as a function of $\mu$. The latter is due to the different orientation of band inversion when changing Rashba parameters from $|\lambda_1| > |\lambda_2|$ to $|\lambda_1| < |\lambda_2|$, as discussed above, with respect to the current direction. Therefore, asymmetric relaxation processes would be considerable for the y component of spin when band inversion is perpendicular to the current direction. 
On the other hand, for the case where the band inversion is parallel with the current direction, $\chi_{yy}$ vanishes. Moreover, at a given energy level outside the Lifshitz points while the values of $\chi_{yy}$ change sharply versus $\lambda_1$, see Fig. \ref{fig6}(a), the values of such quantity change smoothly versus $\lambda_2$, see Fig. \ref{fig6}(b), around the origin. Furthermore, from both figures one can see that $\chi_{yy}$ is asymmetric as functions of $\lambda_1$ and $\mu$ but symmetric versus $\lambda_2$. Figure \ref{fig6}(c) illustrates $\chi_{yy}$ as function of $\lambda_1$ for different values of $\lambda_2$ with $\mu = 2\times10^{-2}$. As $\lambda_2$ increases in magnitude, the abrupt sign change of $\chi_{yy}$ around origin at first becomes pronounced and then it smears out. Moreover, away from the origin, $\chi_{yy}$ decreases in magnitude with increasing $|\lambda_2|$. As shown in Fig. \ref{fig6}(d), $\chi_{yy}$  is increased for small absolute values of $\lambda_2$ by increasing $\lambda_1$, while for large absolute values of $\lambda_2$, since $\chi_{yy}$ tends to vanish, the change of values of $\lambda_1$ is almost always insignificant.

Comparing density plots of Figs. \ref{fig5} and \ref{fig6}, one can find that although the values of $\chi_{yy}$ usually are some order of magnitudes smaller than those of $\chi_{xy}$ in the most region of space of parameters, but within Lifshitz points for $|\lambda_1| \gg |\lambda_2|$, the values of $\chi_{xy}$ diminish so that $\chi_{yy}$ becomes dominant. However, interestingly, for $|\lambda_1| \ll |\lambda_2|$ within Lifshitz points both types of spin polarizations, i.e., $\chi_{xy}$ and $\chi_{yy}$, vanish. As a result, remarkably, both of the magnitude and direction of spin polarization are tunable by the Rashba parameters in the present case.

\section {Vertex correction}\label{s4}
The goal of this section is to examine the effects of graphene's disorder \cite{SpiSuscVertexCorr,disorGraph1} on current-induced spin polarization in the presence of anisotropic Rashba spin-orbit interaction. It is well-known that the effect of impurities can be taken into account by renormalizing defect-free system via self-energy as well as spin-vertex function. The effect of self-energy is discussed above. The spin-vertex function $\tilde{S}_{\alpha}$ can be obtained by iteration through Eq. (\ref{VereqnkR3}). But, since the considered impurities have pointlike feature, we can decompose the spin-vertex function with the aid of Pauli matrices acting on the spin space \cite{curSpinGraph2} as
\begin{equation}
\tilde{S}_{\alpha} = a_{\alpha} \frac{\hbar}{2}\tau_{0}\otimes\sigma_{x} + b_{\alpha} \frac{\hbar}{2}\tau_{0}\otimes\sigma_{y} + c_{\alpha} \frac{\hbar}{2}\tau_{0}\otimes\sigma_{z} + d_{\alpha} \frac{\hbar}{2}\tau_{0}\otimes\sigma_{0}.
\end{equation}%
Here, the unknown coefficients $a_{\alpha}$, $b_{\alpha}$, $c_{\alpha}$ and $d_{\alpha}$ for $\alpha = x, y, z$ may be easily obtained by using the procedure described in Ref. \cite{curSpinGraph2}. We only notice that for our model $d$'s vanish identically and $c$'s have no significant contribution to the polarization of current which will be disregarded in what follows. Therefore, the non-zero coefficients are given by,
\begin{eqnarray}
a_x &=&\frac{1}{1-n_{imp}V^2_0 \mathcal{I}_x},\\
b_y &=&\frac{1}{1-n_{imp}V^2_0 \mathcal{I}_y}, \label{acofxy}
\end{eqnarray}%
where
\begin{equation}
\mathcal{I}_{x(y)} = \int \frac{dkd\theta}{(2\pi)^2} \frac{f_{x(y)}(\mu,\Gamma)}{\prod_{n=1}^4[(\mu-E_n)^2+\Gamma^2]}. \label{aspxy}
\end{equation}%
Here, the expressions for $f_{x}(\mu,\Gamma)$ and $f_{y}(\mu,\Gamma)$ are a little lengthy even by retaining their terms up to leading-order in $\Gamma$, we therefore refrain from expressing them explicitly. Consequently, a certain in-plane spin state will be scattered on non-magnetic impurities so that the two components of spin polarization will be modified by different numerical factors.
\begin{figure}
 \centering
  \includegraphics[width=8.5cm]{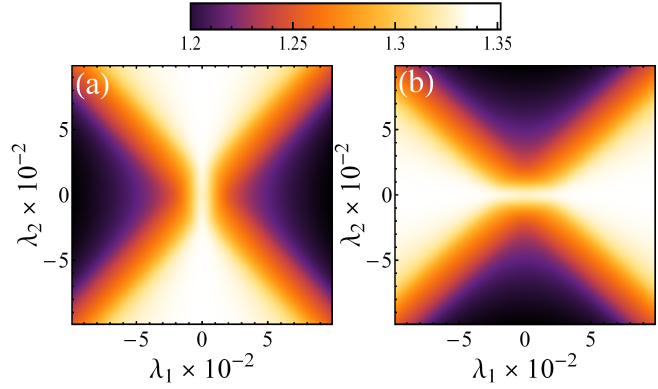}
  \caption{(Color online) Density plot of (a) $a_{x}$ and (b) $b_{y}$ as functions of $\lambda_1$ and $\lambda_2$ with $n_{imp}V^2_0 = 10^{-3}$ and $\mu = 2\times10^{-2}$.}
  \label{fig7}
\end{figure}
The evaluated coefficients $a_{x}$ and $b_{y}$ are illustrated in Fig. \ref{fig7}. From Fig. \ref{fig7} (a) we see that $a_{x}$ increases and then decreases as $\lambda_1$ increases. Whereas for a given $\lambda_1$ with increasing $\lambda_2$, the value of $a_{x}$ decreases and then increases except for values of $\lambda_1$ around the origin. For $b_{y}$, as shown in Fig. \ref{fig7} (b), the behavior is reversed as compared to the case $a_{x}$. Since the modification of $x$ and $y$ components of spin vertex function is different in the presence of anisotropic spin-orbit coupling, therefore, interestingly, scattering from non-magnetic impurities leads to directional dependence of the resulting spin states and the spin polarization.

\section {Summary and final comments} \label{s5}
In summary, we demonstrated the electrically-induced spin polarization of carriers in the presence of anisotropic Rashba spin-orbit coupling in a disordered graphene layer. The Green's function method along with the linear response theory are used to calculate the magnitude and direction of the current-induced spin polarization. We showed analytically and numerically that spin polarization can have in-plane components along and perpendicular to the electric current orientation. While in the case of isotropic Rashba spin-orbit coupling, the dependence of perpendicular component of spin polarization on Fermi energy around the Dirac point is linear, anisotropic Rashba spin-orbit coupling gives rise to the non-linear dependence of spin polarization. This behavior takes place at the range of energies which is within the Lifshitz points of the conduction and valance bands. Furthermore, we argued that anisotropic Rashba parameters have the characteristic feature providing a new way to control spin-polarized current. This property allows for the manipulation of magnitudes and subsequently directions of both spin polarization components. With the inclusion of spin-vertex correction owing to scattering on non-magnetic impurities, we elucidated that both components of the current-induced spin polarization will be modified by numerical factors. These factors are significantly different which this is simply reflecting the fact that Rashba spin-orbit coupling has different weights in different spin directions.

\par
Rashba anisotropy would be induced by either a dilute number of adatoms or intercalated superlattices \cite{AnisoRashGraph1} underneath graphene so that the charge-carrier mobility properties and lattice structure of graphene itself are not strongly affected. Due to band hybridization of the intercalated island or large clusters with graphene, their relative structural arrangement can reduce the symmetry of graphene from the hexagonal symmetry. Such hybridization can transfer the spin-orbit interaction within the cluster bands to graphene bands resulting in non-abelian spin-dependent gauge potential \cite{AnisoRashGraph1}. This gauge potential can be translated into Rashba spin-orbit coupling with different weight. Also, the strength of proximity-induced spin-orbit coupling was experimentally estimated ranging from several meV to a few tens of meV depending on material used for adatoms or clusters \cite{SOpt,AnisoRashGraph1}.

\par
As previously discussed, Rashba anisotropy results in the anisotropic shape of the Fermi contours. Therefore, one may expect that the anisotropic Fermi surface would be responsible for anisotropic charge and heat transport as well as optical conductivity with nontrivial behavior through anisotropic Rashba region. Such effects, however, are of particular interest to investigate and require a detailed analysis in separate papers.

\par
For experimental realization of spin polarization several techniques have been developed using magneto-optical Kerr effect \cite{CurrIndgeneraExper2,CurrIndgeneraExper5}, circularly polarized luminescence \cite{CurrIndgeneraExper3}, and appropriate spin-sensitive electric probes \cite{CurrIndTI2}. In the latter case, using ferromagnetic tunnel barrier surface contacts with different magnetization directions enables one to probe the current spin polarization manifesting itself as a voltage on the ferromagnetic contact. The voltage measured depends on the projection of the spin polarization onto the contact magnetization. The magnetization direction of the ferromagnetic metal can be adjusted by applying a small in-plane magnetic field. Therefore, the possible predicated current-induced spin polarization in the present study can be determined by voltage measurements of ferromagnetic contact that can be magnetized in various orientations.

We should note that an effective low energy Hamiltonian and dc electric field have been used throughout this paper. Therefore, our results would be valid only at low chemical potentials and negligible electric field variation. For high doping and high frequency electric field, full band structure of the material and dynamical spin susceptibility should be taken into account, respectively. Furthermore, external electric field is regarded as a weak perturbation to the system so that our considerations become reliable in the linear response regime.

\ack
The author thanks A. Mohammadi for fruitful discussions. This study was supported by Iran Science Elites Federation under Grant No. 11/66332.


\end{document}